\documentclass[apl,twocolumn,showpacs,superscriptaddress]{revtex4-1}
\usepackage{amsmath,amssymb,graphicx,epstopdf,color,bm,soul}

\begin{document}

\title{Operation of a semiconductor microcavity under electric excitation}

\author{D. V. Karpov}
\affiliation{Institute of Photonics, University of Eastern Finland, P.O. Box 111 Joensuu, FI-80101 Finland}
\affiliation{ITMO University, St. Petersburg 197101, Russia}

\author{I. G. Savenko}
\affiliation{Center for Theoretical Physics of Complex Systems, Institute for Basic Science, Daejeon 34051, Republic of Korea}
\affiliation{Nonlinear Physics Centre, Research School of Physics and Engineering, The Australian National University, Canberra ACT 2601, Australia}

\begin{abstract}
We present a microscopic theory for the description of the bias-controlled operation of an exciton-polariton-based heterostructure, in particular, the polariton laser. Combining together the Poisson equations for the scalar electric potential and Fermi quasi-energies of electrons and holes in a semiconductor heterostructure, the Boltzmann equation for the incoherent excitonic reservoir and the Gross-Pitaevskii equation for the exciton-polariton mean field, we simulate the dynamics of the system minimising the number of free parameters and build a theoretical threshold characteristics: number of particles vs applied bias. This approach, which also accounts for the nonlinear (exciton-exciton) interaction, particle lifetime, and which can, in principle, account for any relaxation mechanisms for the carriers of charge inside the heterostructure or polariton loss, allows to completely describe modern experiments on polariton transport and model devices.
\end{abstract}

\maketitle


Semiconductor microcavities under incoherent background pumping, either electrical or optical, can be used in a variety of applications, such as optical routers~\cite{OurAPL, BlochAPL}, sources of terahertz radiation~\cite{RefTHzAPL, RefTHzPRL}, high-speed optical polarization switches~\cite{Nat4,Wertz2010}. 
In this context, electricaly pumped microcavities have application-oriented perspective, for obvious reasons. Furthermore, such wide-bandgap semiconductors as InAlGaN alloys are promising materials for room-temperature polariton Bose-Einstein condensation (BEC), and thus room-temperature lasing due to large oscillator strength, exciton binding energy, and giant Rabi splitting~\cite{Apl93,Nanolett}.

Bose-Einstein quasi-condensates of exciton polaritons (EPs) form when incoherent electrons, holes and photons scatter their energy, through interaction with other particles, then they couple and form hybrid modes (EPs), and further these eigenmodes of the system collect into a low-energy state~\cite{Kasprzak2006, Balili2007, Lai2007} referred to as the single-particle ground state. 
While conventional Bose-Einstein condensation (BEC) is defined as a macroscopic occupation of the ground state in thermal equilibrium, here one has to deal with a quasi-condensation since the thermal equilibrium in solid state
systems is never achieved due to the finite lifetime of the particles which in the case of EPs amounts to 10-100 ps in modern structures~\cite{SnokePRB882353142013, SnokeOptika212015, SnokearXiv1602030242016,SnokearXiv1601025812016}.

Short lifetime of EPs makes the system highly nonequilibrium~\cite{Wouters2007}, although spatial coherence has been recently reported~\cite{Krizhanovskii2009, Maragkou2010, OurPRL1132039022014}. The theoretical description of such condensates thus requires a kinetic approach, where crucial role is played by the pumping source which should continuously feed the system in order to compensate the losses. The pumping source usually brings excitation to one of the components: either excitons or photons. Theoretical description of the pump is a challenging issue, especially when we speak about the electrical pumping of the system by application of the bias to the heterostructure and launching electric current through~\cite{Nature453, PRB77, APL92, OurNature4973482013}. There have been suggested several approaches aimed at description of the current injection (e.g.~\cite{RefScientificReports,Savenko2011}), however, they operate with phenomenological equations for the carriers of charge, and thus excitons and polaritons.

%
%
%
\begin{figure}[!b]
\includegraphics[width=1.0\linewidth]{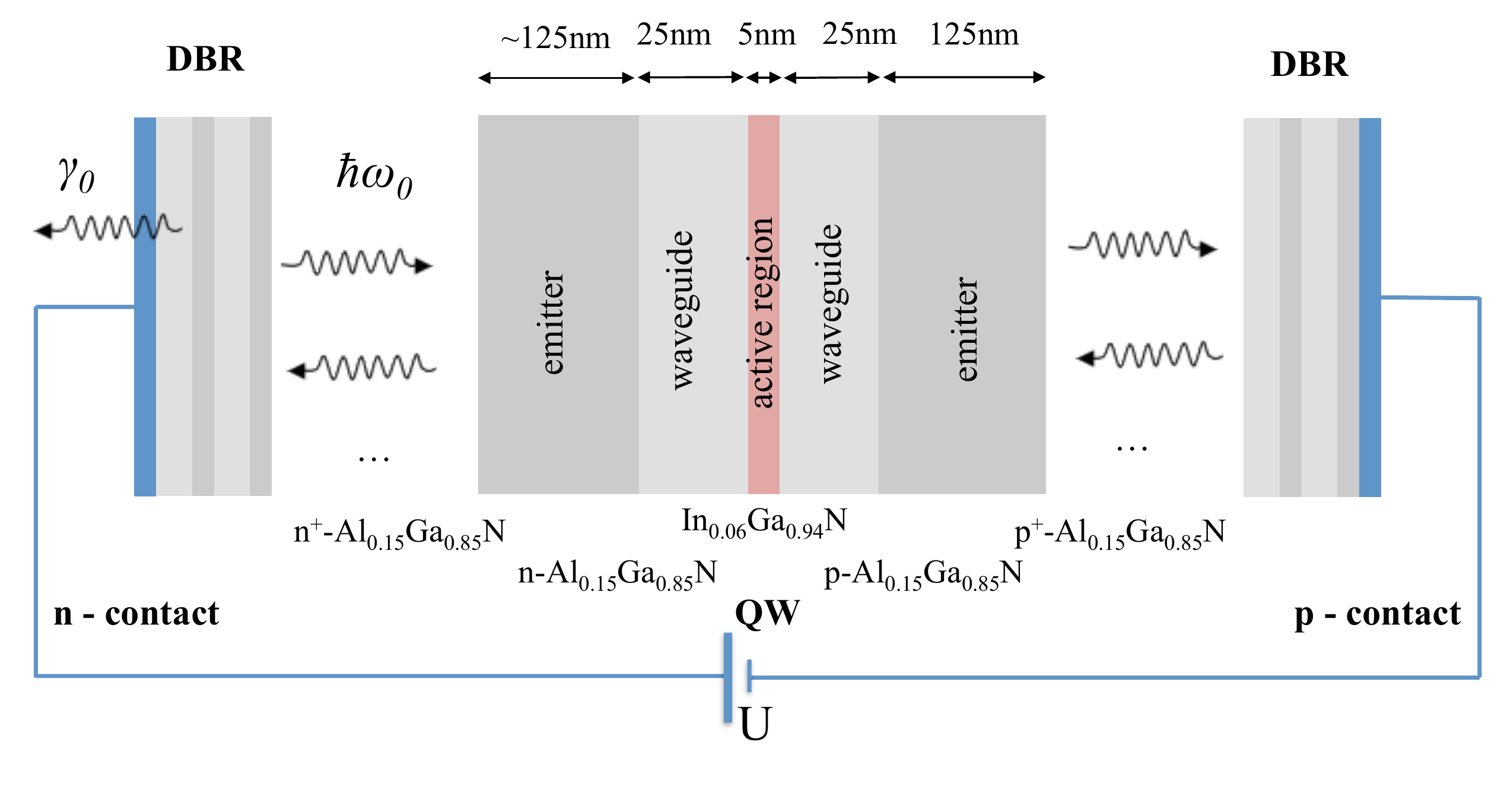}
\caption{Growth stack for InGaN quantum-well (QW) microcavity under electrical excitation. The photons are localised between two Distributed Bragg Reflectors (DBRs) forming a single-mode cavity with frequency $\omega_0$; the excitons are localised in the active region. $\gamma_0$ is the radiative losses rate. Electrical pumping with voltage $U$ is employed to excite the system through bias applied to n-p contacts.} 
\label{FigHetero}
\end{figure}

Interacting EPs can be treated within the Gross-Pitaevskii equation for the mean-fields~\cite{Carusotto2004,OurPRL1101274022013}, which can be modified for incoherent pumping~\cite{Wouters2007,Keeling2008}. Such an approach has been
successful for the description of a variety of recent experiments, including, for example, spatial pattern
formation~\cite{Manni2011,Christmann2012} and spin textures~\cite{SpinRing,Kammann2012}.

In this manuscript we introduce a microscopic theory for the description of electrically pumped polariton laser. In the framework of our formalism, the EP field is coupled to an excitonic reservoir~\cite{Wouters2007} which is, in turn, fed by the electrons and holes in the system. Instead of writing phenomenological kinetic equations for electrons and holes, we write microscopic Poisson-like equations for the Fermi quasi-energies and the scalar electric potential which allows us to build the threshold characteristics.


We consider a microcavity with the growth direction of the heterostricture along the axis $z$ and EPs moving in the $xy$ plane, thus the 3D coordinate is given by $\mathbf{r}=(x,y,z)=(\mathbf{r}_\parallel,z)$. 
For the electric potential, $\phi$, we can write the Poisson equation in the form
\begin{eqnarray}
\label{EqPoisson}
\frac{\partial\phi(\mathbf{r},t)}{\partial t}=-\nabla^2\phi(\mathbf{r},t)-\frac{\rho(\mathbf{r},t)}{\epsilon(\mathbf{r})\epsilon_0},
\end{eqnarray}
where $\epsilon(\mathbf{r})$ is a dielectric permittivity, $\rho=q(N_D^+-N_A^-+p-n)$ is the charge density (here and in the following we omit the explicit notation `$(\mathbf{r},t)$' in $\rho(\mathbf{r},t)$, $n(\mathbf{r},t)$ etc for breivity). $N_D^+$ and $N_A^-$ being ionised donor and acceptor impurity concentrations, 
$N_D^+={N_D}[{1+g_D\mathrm{exp}(\frac{F_n-E_C+E_D+q\phi}{k_BT})}]^{-1}$, $N_A^-={N_A}[{1+g_A\mathrm{exp}(\frac{E_V+E_A-F_p-q\phi}{k_BT})}]^{-1}$
%
%
with $N_D$ and $N_A$ being the full donor and acceptor impurity concentrations; $g_D=2$, $g_A=4$ are the donor and acceptor impurity degeneracy factors, respectively~\cite{RefMorkoc}. In general, $g_A$ may vary from 4 to 6 in conventional nitride semiconductors (due to small splitting of the valence band). 
$E_D$, $E_A$ are the ionization potentials. Further, $E_C$ and $E_V$ are the energies of the conduction band bottom and the valence band top. $F_n=F_n(\mathbf{r},t)$ and $F_p=F_p(\mathbf{r},t)$ are the Fermi quasi-energies of electrons and holes.
$n$ and $p$ are the electron and hole densities. They read the Fermi statistics and are given by
\begin{eqnarray}
\label{EqConcentrations}
n&=&N_C{\cal F}_{1/2}\left(\frac{F_n-E_C+q\phi}{k_BT}\right),\\
\nonumber
p&=&N_V{\cal F}_{1/2}\left(\frac{E_V-F_p-q\phi}{k_BT}\right),
\end{eqnarray}
where $N_C$ and $N_V$ are the densities of states in the Conduction and Valence bands, correspondingly. $N_C=2(m_nk_BT/2\pi\hbar^2)^{3/2}$ with $m_n$ the electron effective mass; and usually $N_V=(m_{lh}k_BT/2\pi\hbar^2)^{3/2}+(m_{hh}k_BT/2\pi\hbar^2)^{3/2}$. However, since polaritons are usually based on the excitons formed of heavy holes, we assume $N_V=(m_{hh}k_BT/2\pi\hbar^2)^{3/2}$ thus neglecting the light hole component. ${\cal F}_{\nu}(\xi)=\Gamma^{-1}(\nu+1)\int_0^\infty x^\nu dx/(1+\mathrm{exp}(x-\xi))$ is the Fermi integral of the order $\nu$, $\Gamma(x)$ is the Gamma-function.
In what follows, we will assume that the electron-hole subsystem of the whole system reaches the steady state much faster than the excitonic and polaritonic susbystems, which is a good approximation in most of real situations. It allows us to consider static electric potential, putting $\partial_t\phi=0$ in \eqref{EqPoisson}.

\begin{figure}[!t]
	\includegraphics[width=1\linewidth ]{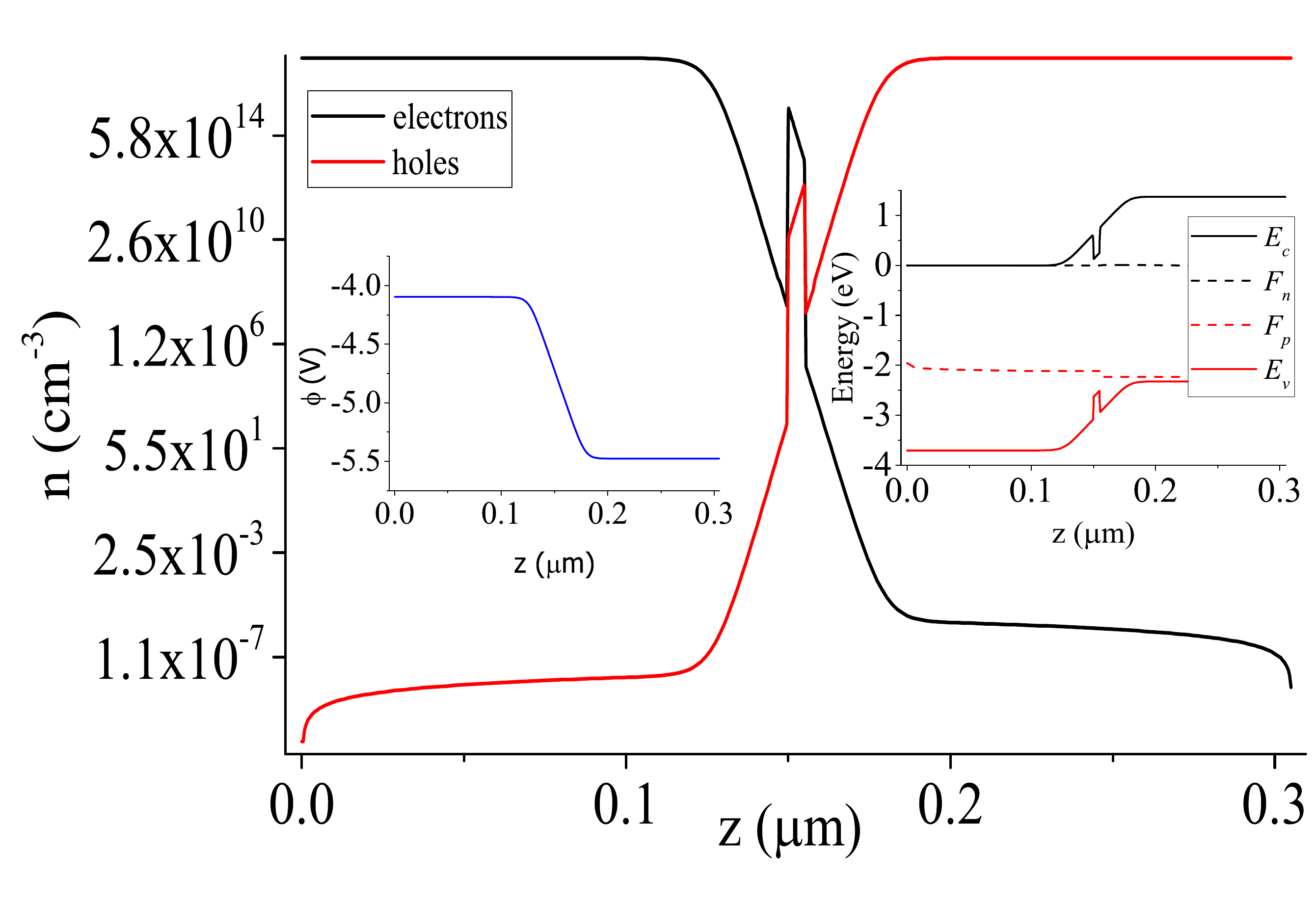}
			\caption{Distribution of the carriers of charge along the heterostructure ($z$-axis) in semi-log scale for the system presented in Fig.~1 under forward bias for the voltages  $U=2.23$ V. Left-hand side inset shows distribution of the scalar potential, $\phi$. Right-hand side inset presents the energy diagram (Conduction and Valence bands energies along $z$ for the corresponding bias.}
	\label{FigDiagrams}
\end{figure}

Now, the key missing ingredient is the spatial distribution of the Fermi quasi-energies. In order to find them, let us write the continuity equations,
\begin{eqnarray}
\label{EqCurrents}
\nabla j_n&=&-q(G-R),~~~~~j_n=\mu_nn\nabla F_n\\
\nonumber
\nabla j_p&=&+q(G-R),~~~~~j_p=\mu_pp\nabla F_p
\end{eqnarray}
where $j_n$ and $j_p$ are the electron and hole current densities, $\mu_n$, $\mu_p$ are the carrier mobilities, $G$ is the carriers generation and $R$ is the general recombination rates, which we take here equal for electrons and holes for simplicity. Using Eq.~\eqref{EqCurrents}, we come up with the Poisson-like equations for the electron and hole Fermi quasi-energies,
\begin{eqnarray}
\label{EqFermi}
\nabla(\mu_nn\nabla F_n)&=&-q(G-R)\\
\nonumber
\nabla(\mu_pp\nabla F_p)&=&+q(G-R).
\end{eqnarray}
Together, Eqs.~\eqref{EqPoisson}, \eqref{EqConcentrations}, and \eqref{EqFermi} represent a closed consistent system of equations and fully describe the electron-hole dynamics with proper boundary conditions. In particular, if we want to simulate the voltage-controlled heterostructure, then for the $z=0$ (n-electrode of the heterostructure) we have $N_D^+-N_A^-+p-n=0$, in the mean time, the bias, $U$ (applied voltage), comes into the equations as
\begin{eqnarray}
\label{EqBiasControl}
F_n(z=0)-F_p(z=L)=qU.
\end{eqnarray}
In our work the only source of pumping is the applied bias, thus we assume $G=0$ in the following. 

The next crucial step is to connect the free charges with the formation of excitons. This we do by the dynamic equations,
\begin{eqnarray}
\label{EqExcitons}
\frac{\partial n_X(\mathbf{r_\parallel},t)}{\partial t}=W~ \tilde{n}~\tilde{p}-\frac{n_X}{\tau_X}-\gamma~ n_X|\psi(\mathbf{r}_\parallel,t)|^2,
\end{eqnarray}
where $n_X$ is the occupation of the reservoir of excitons, $W$ is the rate of exciton formation from the electron-hole plasma, $\tilde{n}$ and $\tilde{p}$ are the densities of electrons and holes which reside in the quantum wells of the heterostructure, and $\gamma$ is the rate of polariton formation fed by the excitonic reservoir. Now we are ready to denote the term R from Eq.~\eqref{EqFermi}, $R=W~\tilde{n}~\tilde{p}$. Thus it accounts for the electron and hole losses due to exchange with the excitonic reservoir. It should be noted, that $R$ can account for various mechanisms of the particle loss. For instance, the non-radiative recombination can be described by the term $\tilde R=\tilde{n}~\tilde{p}(1-\mathrm{exp}[(Fp-Fn)/k_BT])~[\tau_pn+\tau_np]^{-1}$, where $\tau_{n,p}$ are the non-radiative lifetimes of the carriers of charge~\cite{SYuKarpov1}. Besides, the recombination on dislocation cores~\cite{SYuKarpov2} and the Auger recombination can be accounted for.

EPs we describe within the mean field approximation, using the macroscopic wavefunction $\psi(\mathbf{r}_\parallel,t)$ with the
Fourier image $\psi(\mathbf{k}_\parallel,t)$.
\begin{figure}[!t]
	\includegraphics[width=1\linewidth]{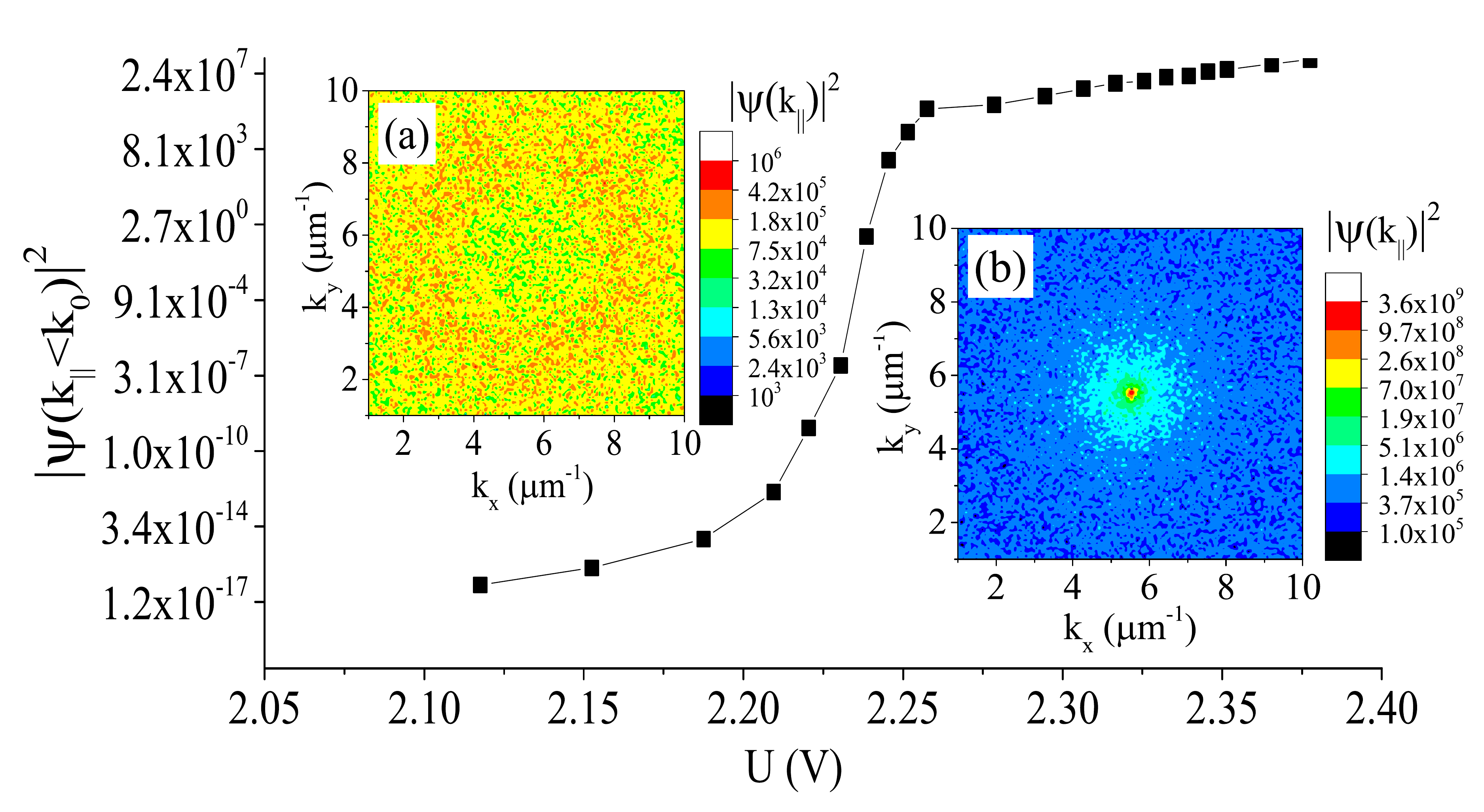}
\caption{Threshold characteristic:  exciton-polariton density in the vicinity of $k_{\parallel}=0$ as a function of forward bias, $U$, for the InGaN quantum-well diode presented in Fig.~1 (see also Fig.~2 for the corresponding distributions of the carriers of charge along $z$). The Bose-Einstein condensation starts at around $U=2.23$ V in $k_0$ vicinity around 0 in k-space (in our modelling we choose $k_0$=2 $\mu m^{-1}$). On the bottom panels, the colormaps of the particle distribution in momentum space for different voltages are presented (a) $U=2.2$ V (under threshold) and (b) $U=2.3$ V (above threshold).} 
	\label{FigThresh}
\end{figure}
The equation of motion reads
\begin{align}
\label{eq:dpsixdt}
i\hbar\frac{d\psi(\mathbf{r}_\parallel,t)}{dt}&={\cal F}^{-1}\left[E_{k_\parallel}\psi(\mathbf{k}_\parallel,t)\right]
+i\frac{\hbar\gamma}{2} n_X(\mathbf{r}_\parallel,t)\psi(\mathbf{r}_\parallel,t)
\notag\\
&+\left[V(\mathbf{r}_\parallel,t)+\alpha\left|\psi(\mathbf{r}_\parallel,t)\right|^2-\frac{i\hbar}{2\tau}
\right]\psi(\mathbf{r}_\parallel,t),
&\hspace{0mm}
\end{align}
where $E_{k_\parallel}$ is the particle dispersion (which is non-parabolic for EPs); 
$V(\mathbf{r}_\parallel,t)$ is the potential profile, $\alpha$ is a constant
describing the strength of particle-particle interactions. It can be estimated as
\cite{Yamamoto1999}: $\alpha\approx E_ba_B^2/(\Delta x\Delta y)$, where$\Delta y=L_y/N$, $\Delta x=L_x/N$ are the discretisation units, $L_{x,y}$ are the spatial dimensions in $xy$.
We have also introduced the decay term $-i(\hbar/2\tau)\psi$ to account for the radiative decay of
particles~\cite{Carusotto2004}.


We consider an InGaAlN alloy-based microcavity presented in Fig.~1. The active region of the heterostructure consists of 5nm In$_{0.06}$Ga$_{0.94}$N QW. It is located between n-Al$_{0.15}$Ga$_{0.85}$N and p-Al$_{0.15}$Ga$_{0.85}$N highly doped regions, commonly referred to as emitters, and less doped regions, the waveguides. The outer layers of the structure are the distributed Bragg reflectors which provide optical confinement. System is pumped by a direct bias, $U$.
In computations we used $\tau=18$ ps. The exciton-polariton dispersion was calculated using a two oscillator model with cavity photon effective mass $4\times10^{-5}$ of the free electron mass, Rabi splitting $10$ meV and exciton-photon detuning $2.5$ meV at zero in-plane wave vector.

Figure 2 shows the carrier concentrations which correspond to the polariton threshold value of voltage, $U\approx 2.23$. With the increase of voltage, the scalar potential distribution changes (left hand side inset) and the Fermi quasi-energies approach the Conduction and Valence bands in the QW region, correspondingly (right inset). Then, high enough concentrations of electrons and holes lead to sufficient concentration of excitons in the QW region and thus formation of polariton BEC. 

Figure 3 is the manifestation of the threshold characteristics for EPs. EP density around $k_\parallel=0$ increases rapidly above threshold voltage, $U=2.23$ V. The diagrams in Fig. 3 shows (a) below-threshold particles distribution (no condensation occurs) and (b) condensation. 
It should be noted, that our formalism allows to account for various scattering mechanisms for EPs also,
for example, involving hot excitons with large momentum~\cite{Porras2002}. Such hot excitons are usually created in non-resonantly pumped systems~\cite{Wertz2012}. In general, our approach allows a theoretical study of the interplay between both exciton mediated and phonon mediated scattering processes in extended systems~\cite{Tassone1997, Piermarocchi1996, Hartwell2010, Yamamoto1999}. However, we do not consider these processes here since description of scattering processes is not the main goal of this manuscript.

It is also known that one of the key signatures of the polariton BEC is the spontaneous coherence buildup. However, since our manuscript is mostly devoted to the development and introduction of the pumping terms, we use a simple conservative Gross-Pitaevskii treatment to model the polariton dynamics. This treatment assumes complete coherence in the system and does not account for the system-environment interaction, thus in its framework, the coherence buildup cannot be checked. However, one can investigate this issue by adding additional terms in the Gross-Pitaevskii equation, employing such approaches as the Truncated Wigner~\cite{prb79}, or the dissipative Gross-Pitaevskii~\cite{OurPRL1101274022013,ourPRBdiss} equation.


{\it Conclusion.---} We have derived a theory for the description of electrically driven exciton-polariton heterostructures, in particular, the polariton laser. Merging the Poisson equations for the scalar electric potential and the Fermi quasi-energies of electrons and holes in a semiconductor heterostructure, the Boltzmann equation for the incoherent excitonic reservoir and the Gross-Pitaevskii equation for the exciton-polariton mean field, we have simulated the dynamics of the system with the minimal number of free parameters and built the theoretical threshold characteristics of the device.

We thank Dr. Oleg Egorov for useful discussions. We acknowledge support of the Australian Research Council's Discovery Projects funding scheme (project DE160100167), President of Russian Federation (project MK-5903.2016.2), and Dynasty Foundation.


\end{document}